\documentclass[apj]{emulateapj}
\usepackage{apjfonts}
\usepackage[usenames,dvipsnames]{xcolor}
\usepackage{slantsc}
\usepackage[T1]{fontenc}
\definecolor{blue}{rgb}{0,0,1}
\usepackage[hyperfootnotes=false]{hyperref}
\usepackage{CJK}
\usepackage{color}
\usepackage{graphicx}
\hypersetup{
  colorlinks,
  citecolor=blue,
  linkcolor=blue,
  urlcolor=blue}

\slugcomment{}

\shortauthors{Liu et al.}

\begin{document}
\begin{CJK*}{UTF8}{gbsn}
\title{Giant Impact: An Efficient Mechanism for the Devolatilization of Super-Earths}
\author{Shang-Fei Liu (刘尚飞) \altaffilmark{1}, Yasunori Hori \begin{CJK}{UTF8}{bsmi}(堀安範)\end{CJK}\altaffilmark{2,4}, D.N.C. Lin \altaffilmark{2,5,6} and Erik Asphaug \altaffilmark{3}}
\affil{$^1$ Department of Earth and Planetary Sciences, University of California, Santa Cruz, CA 95064, USA; sliu26@ucsc.edu \\
$^2$ Department of Astronomy and Astrophysics, University of California, Santa Cruz, CA 95064, USA; yahori@ucsc.edu, lin@ucolick.org \\
$^3$ School of Earth and Space Exploration, Arizona State University, Tempe, AZ 85287, USA; easphaug@asu.edu \\
$^4$ Astrobiology Center, National Institute of Natural Sciences and National Astronomical Observatory of Japan, Osawa 2-21-1, Mitaka, Tokyo 1818588, Japan \\
$^5$ Kavli Institute for Astronomy \& Astrophysics, Peking University, 
Beijing, China \\
$^6$ Institute for Advanced Studies, Tsinghua University and National Astronomical Observatory of China, Beijing, China }

\submitted{Accepted to ApJ}

\begin{abstract}
{Mini-Neptunes and volatile-poor super-Earths coexist on adjacent orbits 
in proximity to host stars such as Kepler-36 and Kepler-11. Several 
post-formation processes have been proposed for explaining the origin 
of the compositional diversity between neighboring planets: the mass 
loss via stellar XUV irradiation, degassing of accreted material, and 
in-situ accumulation of the disk gas. Close-in planets are also likely 
to experience giant impacts during the advanced stage of planet formation. 
This study examines the possibility of transforming volatile-rich 
super-Earths / mini-Neptunes into volatile-depleted super-Earths 
through giant impacts. We present the results of three-dimensional 
hydrodynamic simulations of giant impacts in the accretionary and 
disruptive regimes. Target planets are modeled with a three-layered 
structure composed of an iron core, silicate mantle and hydrogen/helium 
envelope. In the disruptive case, the giant impact can remove most of 
the H/He atmosphere immediately and homogenize the refractory material 
in the planetary interior. In the accretionary case, the 
planet is able to retain more than half of the original gaseous envelope, while 
a compositional gradient suppresses efficient heat transfer as the planetary 
interior undergoes double-diffusive convection.
After the giant impact, a hot and inflated planet cools and contracts 
slowly. The extended atmosphere enhances the mass loss via both a Parker wind 
induced by thermal pressure and hydrodynamic escape driven by the stellar 
XUV irradiation.  
As a result, the entire gaseous envelope is expected to be lost due to the combination 
of those processes in both cases. Based on our results, 
we propose that Kepler-36b may have been significantly devolatilized by 
giant impacts, while a substantial fraction of Kepler-36c's atmosphere may remain intact. 
Furthermore, the stochastic nature of giant impacts may account
for the observed large dispersion in the mass--radius relationship of close-in super-Earths and 
mini-Neptunes (at least to some extent).
}
\end{abstract}


\section{Introduction}

Data obtained from the \textit{Kepler} mission \citep{2010Sci...327..977B} 
indicate an abundant population of close-in planets with radii of 
$R_{\rm p}=$ 1 -- 4\,$R_\oplus$  \citep{2013ApJ...770...69P, Burke:2014dw,
2015arXiv150101623D}. Although the masses of only a handful of those 
planets have been measured with radial-velocity or transit-timing 
variations (TTV) observations, a large range of their mean densities 
$\rho_{\rm p}$ indicates that they have diverse bulk compositions \citep[e.g. 
see Fig.4 in][]{2015ApJ...800..135D}. These kinematic and structural
properties impose constraints on theories of planetary origin. 

It has been suggested that these planets may have formed through the 
coagulation of grains and planetesimals at their present-day location
\citep{2012ApJ...751..158H, 2013MNRAS.431.3444C}. Although all 
volatile grains in their natal disks are sublimated in the proximity
of their central stars, embryos may have formed prolifically from 
refractory grains \citep{Li:2015} and acquired or retained a modest 
amount of hydrogen/helium atmosphere {\it in situ} with relatively 
low apparent mean densities \citep{2012ApJ...753...66I}.  

Close-in planets are exposed to intense XUV (extreme UV and X-ray) 
irradiation from their host stars. Photoevaporation can significantly modify 
the structure of their atmosphere \citep[e.g.,][]{2013ApJ...775..105O}. 
One possible cause for the apparent dispersion in $\rho_{\rm p}$ is the 
retention efficiency of H/He atmospheres among planets around host stars 
\citep{2013ApJ...770..131L} with different XUV luminosities during 
their pre-main sequence stage.  But around some host stars, high-density 
super-Earths and low-density hot-Neptunes with similar $M_{\rm p}$ and $P$ 
(where $M_{\rm p}$ is the planetary mass and $P$ is the orbital period) coexist. 
Their $\rho_{\rm p}$ diversity cannot be simply attributed to the XUV flux from 
their common host stars.

Around a particular sub-giant star, two planets, Kepler-36b and c, were found 
\citep{2012Sci...337..556C} with adjacent orbits ($P = 13.8$ and $16.2$ days),
comparable masses ($M_{\rm p} =4.46 \pm 0.3 \;M_\oplus$ and $8.10 \pm 0.53 
\;M_\oplus$), and radii ($R_{\rm p}=1.48 \pm 0.03 \;R_\oplus$ and $3.67 \pm 0.05 
\;R_\oplus$).  The inferred $\rho_{\rm p}$ of the inner super-Earth (Kepler-36b) 
matches that of an Earth-like composition and that of the outer 
mini-Neptune (Kepler-36c) is consistent with an internal composition 
of a thick hydrogen/helium (H/He) envelope ($8.6\pm1.3$ wt.\,\%, assuming 
an Earth-like core) atop a rocky/iron core \citep{2013ApJ...776....2L}. 
Although a fraction of Kepler 36c's atmosphere may be due to the 
degassing of accreted planetesimals \citep{2008ApJ...685.1237E} 
or \textit{in-situ} the accumulation of a residual disk gas onto its 
solid core \citep{2012ApJ...753...66I}. However, such a large density 
contrast between these two closely-spaced planets is incompatible 
with either of these two accretion processes. 

It is possible that these two planets may once have had a similar amount 
of atmospheric H/He and their compositional dichotomy could be caused by 
their different core masses \citep{2013ApJ...776....2L}. Provided that the 
lost mass is a small fraction of $M_{\rm p}$, their near 7:6 mean motion 
resonance (MMR) orbits would not be significantly affected by it 
\citep{2015arXiv150401680T}.  However, Kepler 36c appears to have
retained its atmosphere and Kepler 36b appears to have not, albeit the former has 
a smaller surface escape velocity and is expected to be more 
vulnerable to photoevaporation than the latter.  Another system with 
a similar dichotomy is Kepler 11b and c \citep{2011Natur.470...53L}.

Another scenario is that protoplanetary embryos formed throughout the disk, 
including the volatile-rich cold outer regions, and converged through 
type I migration to their present-day locations in the proximity of their 
host stars \citep{2010MNRAS.405..573P, 2013ApJ...775...42I}. According to the 
conventional accretion scenario, protoplanetary embryos emerge through 
oligarchic growth and experience cohesive collisions with embryos with comparable 
masses \citep{1998Icar..131..171K}. The energy released in such catastrophic events
intensely heats both the mantle and atmosphere and may induce substantial 
losses of planetary atmosphere \citep{2014ApJ...795L..15S}. 
The giant impact hypothesis is consistent with the chaotic dynamics exhibited in the Kepler-36 system \citep{2012ApJ...755L..21D}.
Hydrodynamic and \textit{N}-body simulations have revealed that stochastic/convergent migration can lead to close encounters between the two planets and physical collisions with other embryos \citep{2013MNRAS.434.3018P,2013MNRAS.435.2256Q}.

Due to the stochastic nature of these giant impact events, structural diversity among 
members of closely packed, multiple short-period planet systems is the 
hallmark of oligarchic growth, extensive migration, and giant impacts. 
This diversity takes the form of different levels of shock processing, 
different final material inventories and planetary densities \citep{2010ChEG...70..199A}.

In this paper we explore the possibility that structural diversity
is established by giant impacts during the formation of compact multiple-planet systems.
Collisions and disruptions of differentiated super-Earths with terrestrial 
compositions \citep{2009ApJ...700L.118M} or water : rock ratios of $1 : 1$ 
\citep{2010ApJ...719L..45M} have been previously simulated with the smoothed 
particle hydrodynamics (SPH) scheme \citep{2012Icar..221..296R}. 
However, there have been no systematic 
simulations of giant impacts between super-Earths with H/He atmospheres. 

In \S2, we outline the numerical method and the initial and boundary conditions 
used for these simulations.  In our models, we adopt physical parameters
similar to those of the Kepler 36 system and scrutinize the collisional 
origin of the compositional diversity between Kepler-36b and c. In \S3,
we present the results of two sets of numerical simulations to show that 
energetic giant impacts can cause strong shocks and lead to the ejection 
of a large amount of the H/He envelope. Shock dissipation also heats up the 
planetary atmosphere and interior, elevating the mass-loss rate induced 
by hydrodynamic escape and radiative evaporation \citep{2004ASPC..321..203H}. 
We summarize our results and discuss their implications in \S4.

\section{Computational method}

\begin{deluxetable*}{cccccccccccccccccc}
    \tabletypesize{\scriptsize}
    \tablecolumns{18}
    \tablewidth{0pt}
    \tablecaption{Mass and energy budgets for the low and high speed impacts. \tablenotemark{a}\label{tab:t1}}
    \tablehead{\colhead{$v_{\rm imp}$} 		&
		\multicolumn{3}{c}{$M_{\rm T}$}	&
		\multicolumn{2}{c}{$M_{\rm I}$}		&
		\multicolumn{3}{c}{$M$}	&
		\colhead{$U_{\rm T}$}	&
		\colhead{$E_{\rm k, T}$}	&
		\colhead{$U_{\rm I}$}	&
		\colhead{$E_{\rm k, I}$}	&
		\multicolumn{2}{c}{$U$}			&
		\multicolumn{2}{c}{$E_{\rm k}$}		\\
		\colhead{\scriptsize (km/s)}		&
		\colhead{\scriptsize Iron\tablenotemark{b}}		&
		\colhead{\scriptsize Rock\tablenotemark{c}}		&
		\colhead{\scriptsize H/He\tablenotemark{d}}		&
		\colhead{\scriptsize Iron}		&
		\colhead{\scriptsize Rock}		&
		\colhead{\scriptsize Iron}		&
		\colhead{\scriptsize Rock}		&
		\colhead{\scriptsize H/He}		&
		\colhead{}			&
		\colhead{}			&
		\colhead{}			&
		\colhead{}			&
		\colhead{$r_{\rm P}$\tablenotemark{e}}		&
		\colhead{$r_{\rm H}$\tablenotemark{f}}		&
		\colhead{$r_{\rm P}$}		&
		\colhead{$r_{\rm H}$}		
		}
    \startdata
     10.96  & 1.344 & 2.663 & 0.297 & 0.334 & 0.663 & 1.676 & 3.315 & 0.208 & 4.746 & 0.441 &  0.516 & 2.092 & 9.176 & 9.442 & 0.186 & 0.215 \\
     53.06  & 3.115 & 6.129 & 0.746 & 0.334 & 0.662 & 3.449 & 6.289 & 0.151 & 21.890 & 4.529 & 0.407 & 72.338 & 31.039 & 32.810 & 0.240 & 0.311
    \enddata
    \tablenotetext{a}{Integral quantities $M$, $U$ and $E_{\rm k}$ 
correspond to mass, internal energy and kinetic energy, respectively. 
Subscript T and I denote the target and the impactor. Integral quantities 
without additional subscripts are measured at the end point of our 
simulations. Mass data are in units of Earth mass and energy data are 
in units of $10^{39}$ erg.}
    \tablenotetext{b, c, d}{Mass of species iron, rock and H/He, respectively.}
    \tablenotetext{e}{The integral quantity is measured within the original radius of the target (cf. Figure \ref{fig:f2} and \ref{fig:f7}).}
    \tablenotetext{f}{The integral quantity is measured within the Hill radius ($2.62 \times 10^{10}$ and $3.21 \times 10^{10}$ cm, respectively).}
\end{deluxetable*}

Our three-dimensional hydrodynamic simulations are based on the framework 
of FLASH \citep{2000ApJS..131..273F}, an Eulerian code with an adaptive-mesh 
refinement (AMR) capability. This finite-volume scheme has advantages 
over most commonly-used SPH codes in terms of capturing shock waves as well 
as characterizing low-density and sparse regions. 
Besides, advection of different fluid species in a grid cell is allowed to obey 
their own advection equations, which is ideally suited for mixing problems.
The main purpose of this work is to investigate the general aftermath of giant impacts, 
so our study here is on head-on collisions.

In order to minimize the artifact of artificial diffusivity 
\citep{2008MNRAS.390.1267T}, collision simulations are performed in a pair 
of planets' center-of-mass frame rather than in the inertial reference 
frame in which they move through the numerical grids with a Keplerian speed.
The width of the computational domain is $1\times 10^{13}$ cm 
on each side, and we choose an open boundary condition for our simulations.
A multiple expansion with a large angular number ($L_{\rm max}=60$) 
about the center of mass of the embryo pair is adopted in order 
to compute the self-gravity with a sufficient angular resolution. 

Most hydrodynamic simulations of giant impacts ignore the presence 
of a central star and its tidal force. In the present context, the 
Hill radius of a planet, $r_{\rm H} = a\; \left(q/3\right)^{1/3}$ 
(where $a$ is the semi-major axis of the planet and $q$ is the mass 
ratio of the planet to its parent star) is only an order of 
magnitude larger than its physical size. Therefore, the tidal force 
of the central star may have a significant effect on the planet's 
post-impact evolution.

The orbit of the embryo pair around the host star is determined by 
solving a two-body problem that treats a central star and the embryo
pair as two separate point masses \citep{2009ApJ...705..844G}. 
Here we consider giant impacts between a target planet 
$M_{\rm T}$ and an Earth-mass impactor $M_{\rm I}$ at 0.1 AU 
separation from a 1 $M_\sun$ star. Initially, the target and impactor are held apart 
from their center of mass by a few target's radii until they have relaxed 
into a hydrostatic equilibrium subjected to each other's tidal perturbation.  
They are released with their relative velocity gradually increased from 
rest to the values listed in Table \ref{tab:t1}. Further acceleration
is determined by their mutual gravity.

Two simulations of head-on 
collisions in the accretionary and disruptive regimes are performed. 
In the accretionary model 1, $M_{\rm T} = 4.3 \; M_\oplus$, 
i.e. close to the mass of Kepler-36b, and the impact velocity 
$v_{\rm imp} = v_{\rm esc}$, where 
$v_{\rm esc}=\sqrt{2G(M_{\rm T}+M_{\rm I})/(R_{\rm T}+R_{\rm I})}$ 
is the two-body escape velocity. In the disruptive model 2, $M_{\rm T} 
= 10 \; M_\oplus$, i.e. the upper end of the mass spectrum 
that is commonly refer to super-Earths, and the impact velocity $v_{\rm imp} = 3v_{\rm esc}$ 
comparable to the local Keplerian speed. Our model 2 represents the largest scale of 
collisions that could happen to a close-in super-Earth with an atmosphere. Impacts onto a less massive 
super-Earth at the same speed lead to a more disruptive outcome.

In our models, the target is initially composed 
of an iron core, a silicate mantle, and an H/He gaseous envelope (7.5 wt\,\%), 
while the impactor has the same layered structure except without
an atmosphere. The mass ratio of an iron core to a silicate mantle 
is assumed to be $1 : 2$ for all planets. We summarize the masses 
of each species in the targets and impactors in 
Table \ref{tab:t1}. The internal structure of all the 
target planets are resolved by a numerical mesh with $R_{\rm T} / 
\Delta d > 100$, where $R_{\rm T}$ is the target's initial radius and 
$\Delta d$ is the width of the smallest grid cells.

\begin{figure*}
  \centering
  \includegraphics[width= 0.7\linewidth,clip=true]{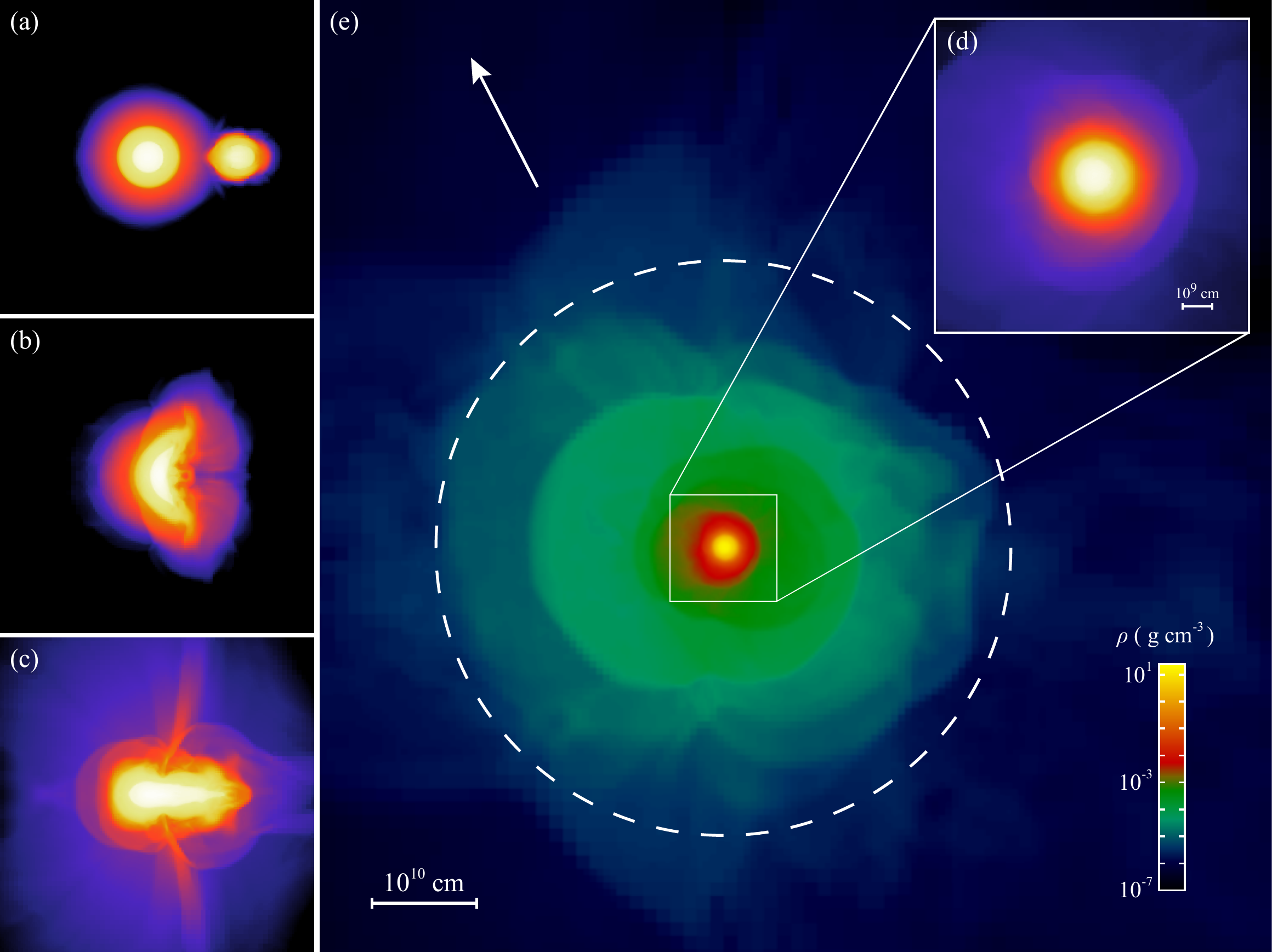}
\caption{Snapshots of the giant impact simulation between a 4.3 
$M_\oplus$ super-Earth and an Earth-mass planet at the escape 
velocity $v_{\rm esc}$. The density distribution across slices 
in the planets' orbital plane with density range [$10^{-4}$ 
, $2.5 \times 10^1$] g cm$^{-3}$ 
(represented by colors ranging from purple to white) before 
the impact (panel a), immediately after (b), at 1.56 hours (c), 
and at 18 hours after the impact (d).  Panel e shows an enlarged 
view of the panel d. With a radius $2.62 \times 10^{10}\,{\rm cm}$,
the dashed circle overplotted on the planet indicates its Hill sphere 
at a distance of 0.1 AU from a solar-mass star. The parent star 
is in the direction of the white arrow.
}
\label{fig:f1}
\end{figure*}

Previously, \citet{2015MNRAS.446.1685L} have applied an early version of this scheme to 
investigate embryo impacts on gas giants.  In those simulations, we 
approximated the internal structure of gas giants with a composite 
polytropic equation of state \citep[EOS; ][]{2013ApJ...762...37L}. 
We assume that the multi-species fluid obeys the Dalton's law, i.e. the total pressure of 
a mixture of gas equals the sum of the contributions of individual components. The current 
study focuses on collisions between smaller planets with compositions and 
internal structures similar to those of the Earth or Neptune.  For 
'terrestrial' types of planetary materials, we have 
incorporated the Tillotson EOS to model iron cores 
and silicate mantles \citep[see Appendix I of][]{Melosh:1989uq}. 
We additionally model an H/He atmosphere (70 wt.\,\% H$_2$ 
and 30 wt.\,\% He) assuming a polytropic EOS ($P \propto \rho^\gamma$) with an
adiabatic index of $\gamma = 5/3$. This simplified assumption can be 
validated for temperatures below the excitation temperature of the
${\rm H}_2$ rotation band, 170K, and above ${\rm H}_2$ dissociation 
temperature ($\sim 2000\,{\rm K}$) but below that for the H and He 
atoms to be fully ionized.  For the intermediate regime, our treatment 
underestimates the compressibility and heat capacity of hydrogen, which 
may exaggerate an impact-driven atmospheric escape. However, the 
atmosphere of the target is generally not sufficiently massive to stall 
the impactor and to prevent the dissipation of its kinetic energy near 
the target's core. We find that the computed
turbulent mixing of material and the mass loss rate of the atmosphere 
are relatively insensitive to the specific H/He EOS model.

Giant impacts are violent and impulsive events. A new quasi-hydrostatic 
equilibrium within the planet's Hill's radius is generally established 
within a dynamical timescale (within a day). During the transitory phase, 
we can neglect the effect of stellar heating on the inflated atmosphere
(see \S4).  After the impact, the efficiency of insolation due to stellar 
irradiation on the planet's surface and the efficiency of heat transfer 
in its interior determine the mass-loss rate from the inflated planet 
\citep{2015arXiv150602049O}. Here we must also consider that the giant 
impact has left behind a massive debris torus that can remain opaque for some
time.

Part of the analysis and visualization presented in Section \ref{sec:s3} is generated using the YT package \citep{2011ApJS..192....9T} and the VisIt software \citep{HPV:VisIt}.

\begin{figure}
  \centering
  \includegraphics[width=\linewidth,clip=true]{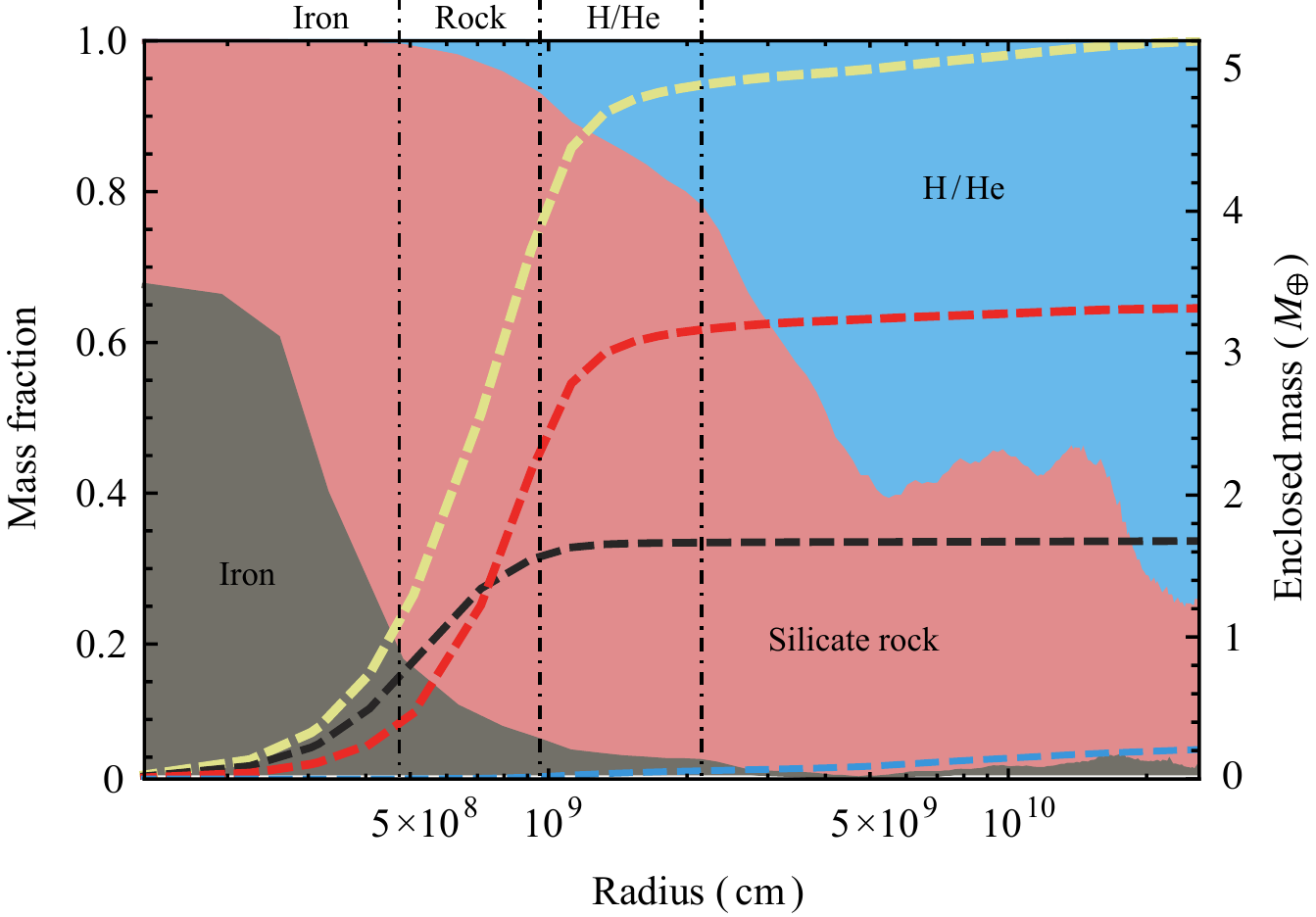}
\caption{Each shaded area illustrates the mass fraction of each species 
as a function of radius at 18 hours after a low-speed giant impact.
The grey, light coral, and blue regions represent iron, silicate rock and 
H/He gas, respectively.  Three vertical dot-dashed lines indicate
the initial compositional boundaries inside the target that has 
the three-layered structure.  The dashed lines from top to bottom 
show the cumulative mass (yellow), the enclosed mass of silicate 
rock (red), iron (black), and H/He (blue) as a function of radius 
in units of Earth mass.}
\label{fig:f2}
\end{figure}

\begin{figure}
  \centering
  \includegraphics[width=\linewidth,clip=true]{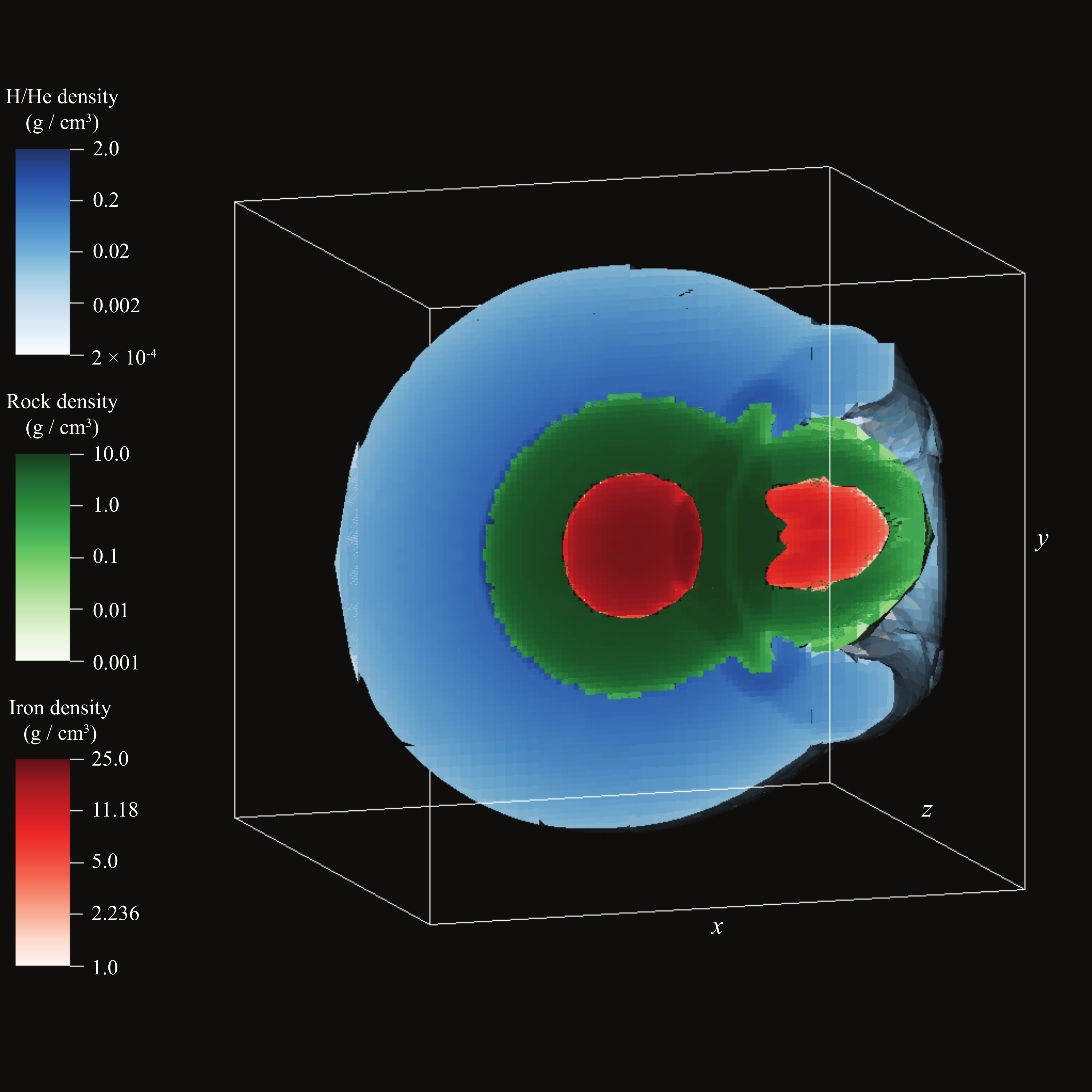}
\caption{Cutaway view showing a snapshot after the contact between the rocky layers of the two planets and before the merger of their iron cores. The blue, green and red colors represent partial densities of H/He, rock and iron species, whichever is the dominant species of a grid cell. For the most part of the target, the total density of a grid cell is close to the partial density of the dominant species because mixing is not severe. However, a small fraction of H/He (less than 5 wt. \%) being mixed with rock is responsible for the formation of a low-density arced structure between compressed rocky layers. The cubic box has a width of 4$\times 10^9$ cm.}
\label{fig:f3}
\end{figure}

\section{Results} 
\label{sec:s3}
\subsection{Low-speed model 1}
The compositional structure of a super-Earth after a low-speed collision
with an Earth-mass impactor is shown in Figure \ref{fig:f1}.  The initial
mass of the target is 4.3 $M_\oplus$. The impactor's approaching 
speed is the escape velocity of the target-impactor system.  
Panels (a) -- (c) show snapshots of density contours before, 
during, and at 1.56 hr after the impact. The snapshots in 
panels (d) and (e) are taken at 18 hours after the impact.
At this time, the central region of the post-impact target is mostly 
dynamically relaxed (see panel (d)), while the hot atmosphere
extends well beyond the Hill radius and continues to lose its mass 
via Roche lobe overflow  (see also \S 4).

Since the total orbital energy of the target-impactor system is nearly zero, 
this impact has an accretion efficiency close to but not quite one. 
The target gains mass after the impact (Table \ref{tab:t1}) but not all of it.
The excess kinetic energy 
injected by the projectile drives an outflow as the planet approaches 
a hydrostatic equilibrium. As a result, the impact is not a perfect merger.
The total mass loss (i.e. the mass that is failed to be accreted by 
the target) at 18 hours after the impact is $\sim 0.1\,M_\oplus$.  
This value is expected to increase slightly due to the slowly decaying 
oscillations in the tenuous outer part of the atmosphere over a prolonged 
period of time. Accretion efficiency in low-velocity collisions can be 
substantially lower than it is for the head-on cases considered here, because 
off-axis giant impacts can attempt to accrete too much
angular momentum for a single planet or planet-satellite system to sustain.
The so-called hit-and-run regime is relevant at velocities intermediate 
to those considered here \citep{2010ChEG...70..199A}.

The gravitationally unbound mass, $\sim 0.1\,M_\oplus$, is mostly composed 
of H/He gas plus a minute contribution from iron and silicate material 
(see Table \ref{tab:t1}).  The impact leads to the immediate ejection
of nearly one-third of the target's initial H/He atmosphere. Most
of the impactor's mass is added to that of the target. Consequently,
the volatile fraction of the target is significantly reduced.

The enclosed mass as a function of radius inside the Hill sphere is
plotted with a yellow dashed line in Figure \ref{fig:f2}. 
Most of the planetary mass is confined within its original radius,
while the tenuous atmosphere spreads out by an order of magnitude, 
filling up the entire Hill sphere (see panel (e) of Figure \ref{fig:f1}). 
The grey, pink, and blue regions in Figure \ref{fig:f2} illustrate 
iron, rock and H/He mass fractions as 
functions of planet's radius.
The masses of the three components enclosed within 
a given spherical radius are overplotted with red (rock), 
black (iron), and blue (H/He) dashed lines, respectively.

\begin{figure*}
  \centering
  \includegraphics[width= \linewidth,clip=true]{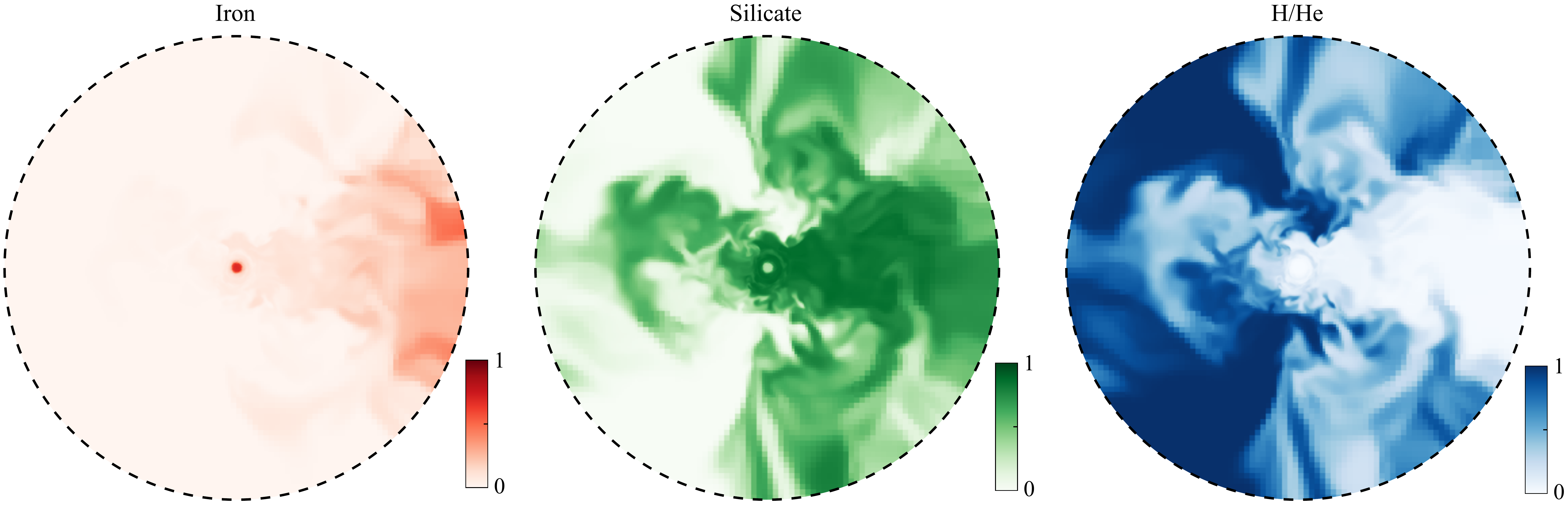}
\caption{Mass fraction slices of each species through the orbital plane at 18 hr after the low-speed impact. A significant degree of mixing is established due to the turbulent mixing as well as hydrodynamic instabilities during the post-impact expansion phase. Dashed circles represent the the Hill sphere with a radius of $2.62 \times 10^{10}\,{\rm cm}$.
}
\label{fig:f4}
\end{figure*}

When the intruding planet enters the target's envelope, a velocity shear is present at the interface between two species which triggers Kelvin-Helmholtz (K-H) instabilities. On the other hand, during the propagation of the impact-induced shock wave through the target, a denser fluid is accelerated by a less dense one at their interface and Rayleigh-Taylor (R-T) instabilities start to develop. However, since the impactor merges with the target in less than half an hour, only small-scale instabilities arise during such a short time span \citep[see, e.g.][]{2007MNRAS.380..963A}, which can only be resolved with an extremely high resolution. Figure \ref{fig:f3} shows a cutaway view of a snapshot when the impactor is about to merge with the target. We plot the partial density, i.e. the density times the mass fraction of the dominant species of a grid cell, to illustrate the interior structure. We note that only a little mixing occurs when the target's gaseous envelope gets crushed to its rocky layer, reducing the density of rocky material in a narrow region between the target and the impactor (see Figure \ref{fig:f3}). As during the short impact phase, both K-H and R-T instabilities are unable to cause large scale mixing, we identify that the impact-driven turbulence is responsible for the global mixing observed during the late stages (see Figure \ref{fig:f4}). Because a large amount of kinetic energy is released upon the coalescence of the target and the impactor, fluid motions become turbulent and a complex mixture of K-H and R-T instabilities can be triggered. 
The turbulent mixing, as well as hydrodynamic instabilities, destroy the original layered interior structure of the target
 -- causing radial mixing between iron and rock near the center,
and inducing a fraction of rock (and a much smaller fraction of iron) 
to diffuse into the H/He atmosphere (see Figure \ref{fig:f4}). Nevertheless, figure \ref{fig:f2} shows that more than two-thirds 
of the mass at the center still consists of iron species.
Besides, the iron mass fraction falls off rapidly with the
radius, indicating that the planetary core survives from the 
impact and its mass grows in a coalescent manner. 

\begin{figure}
  \centering
  \includegraphics[width=\linewidth,clip=true]{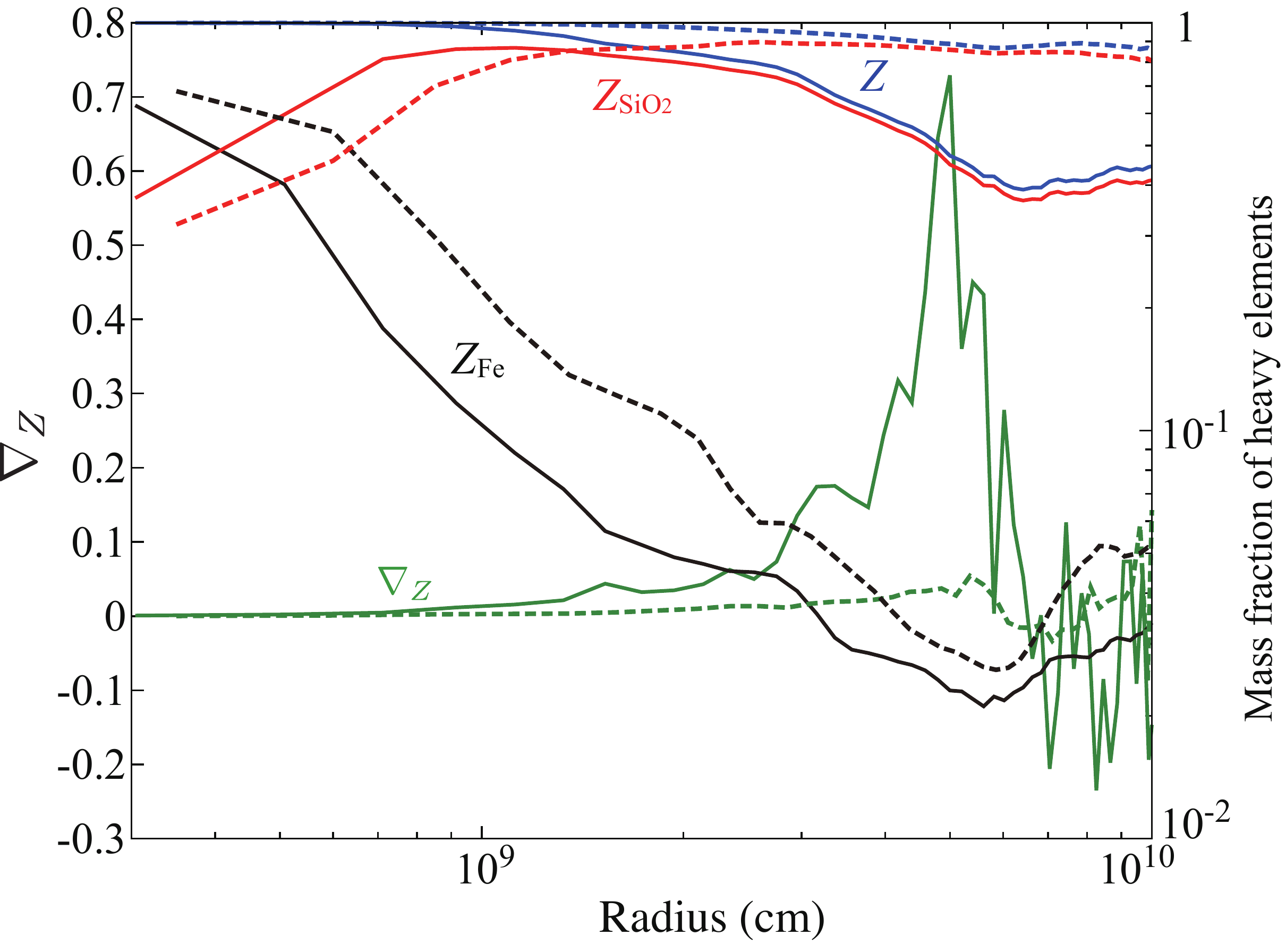}
\caption{Compositional gradients inside the target planets 
after head-on collisions. The low-speed and high-speed 
impacts are represented by solid and dashed lines, respectively. 
Mass fractions of iron (black), rocky material (red), and 
heavy elements (blue) as functions of the radius are shown with
labels on the right. The compositional gradient $\nabla_Z$ 
(green) increases with radius in the low-speed case.  The
small value of $\nabla_Z$ at radii indicates that the elements 
are well-mixed and homogenized. Fluctuation in $\nabla_Z$
reflects that the planet has not yet reached a relaxed state.
}
\label{fig:f5}
\end{figure}

\begin{figure*}
  \centering
  \includegraphics[width= 0.7\linewidth,clip=true]{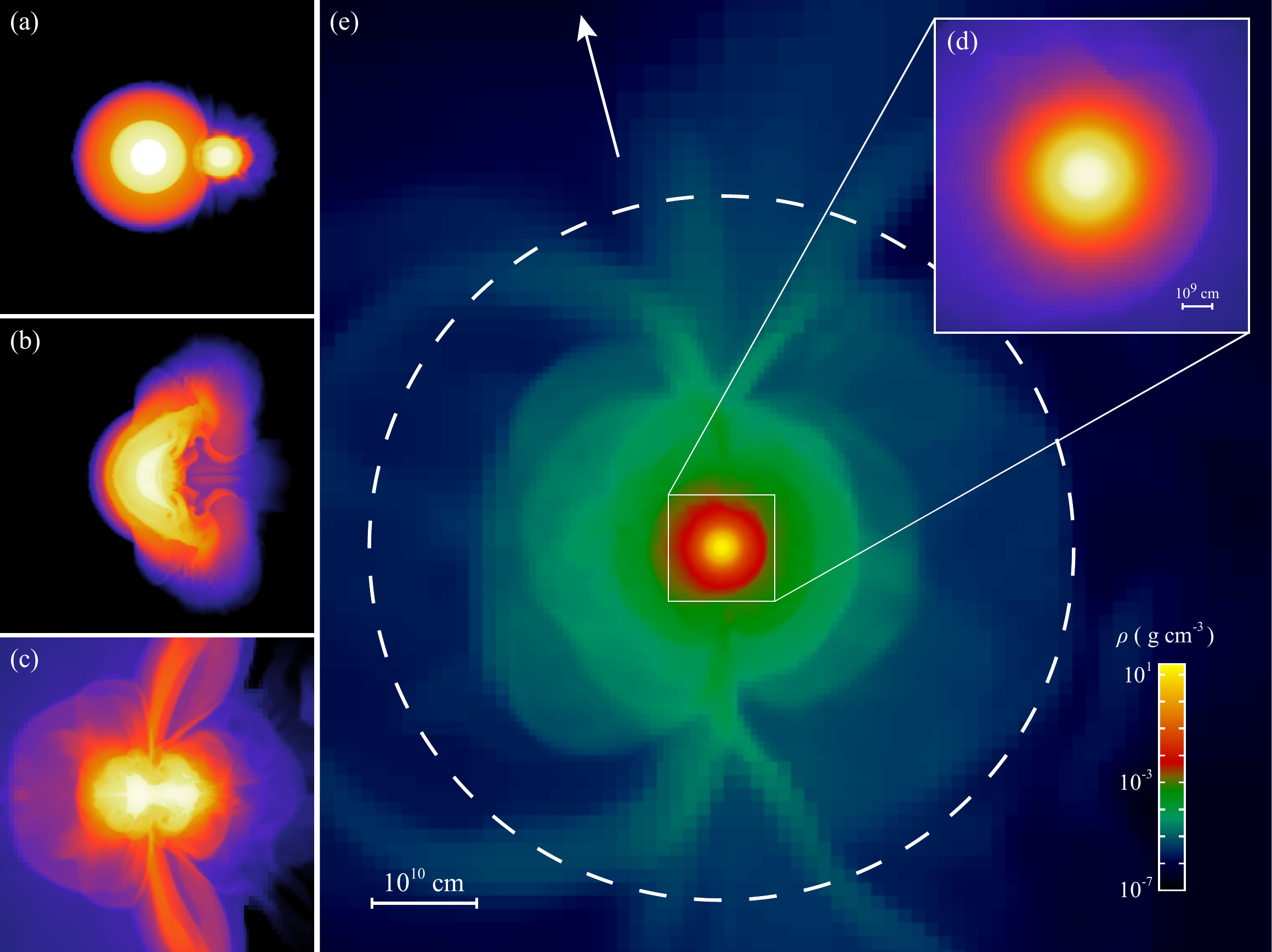}
\caption{Snapshots of the giant impact simulation between a 10 
$M_\oplus$ super-Earth and an Earth-mass planet at
3 $v_{\rm esc}$. The colormaps and symbols are same as in Figure \ref{fig:f1}. 
Panels (a)--(d) are snapshots taken at the start of the impact, at 15 minutes, at 1.5 hr and at 21.5 hr after the impact, respectively. Panel (e) shows an enlarged view of the panel (d). The Hill sphere illustrated by the dashed circle has a radius of $3.21 \times 10^{10}\,{\rm cm}$.
}
\label{fig:f6}
\end{figure*}

\begin{figure}
  \centering
  \includegraphics[width=\linewidth,clip=true]{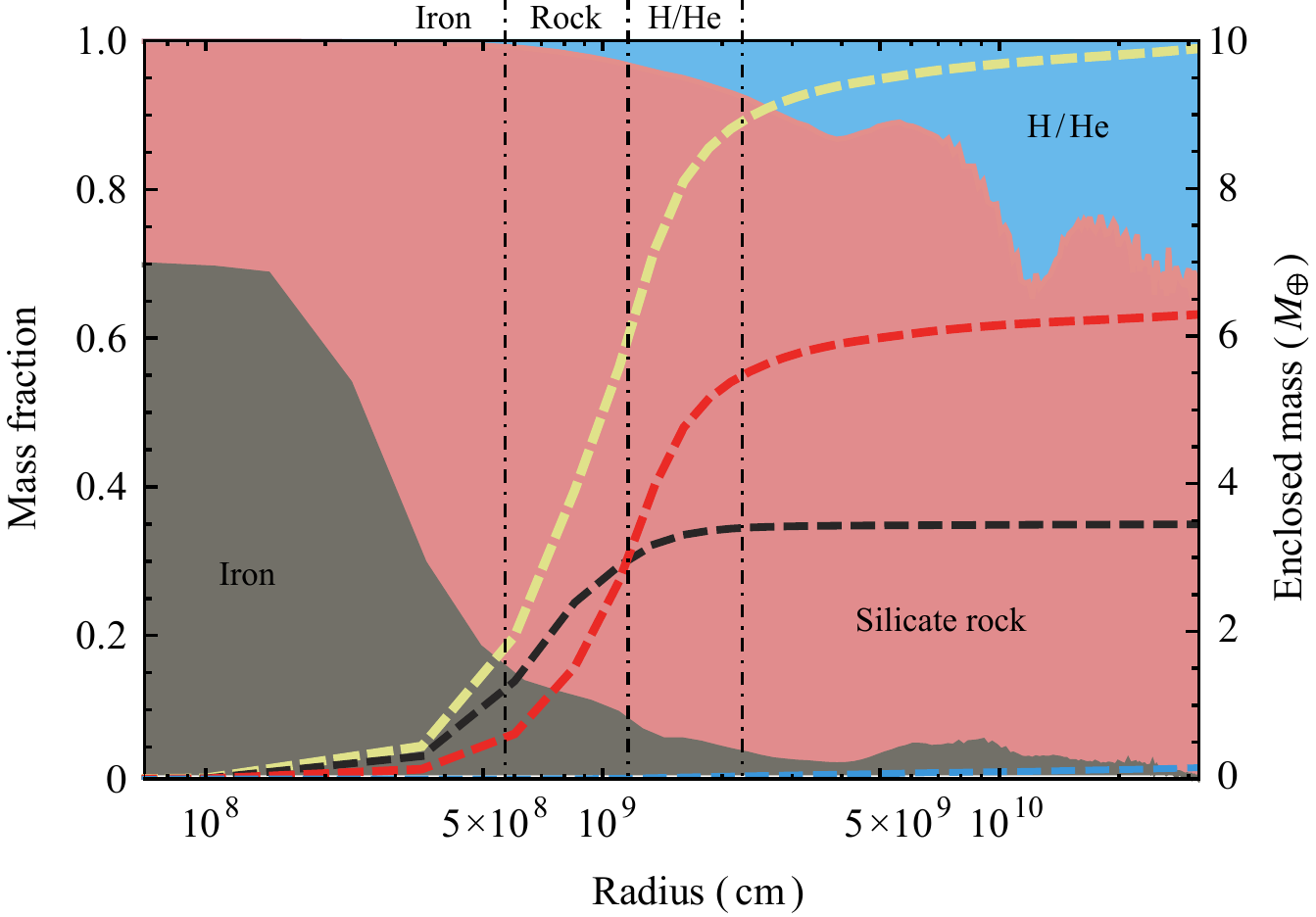}
\caption{Mass fraction area plot and enclosed mass line plot at 21.5 
hours after the high-speed impact. The symbols have the same meanings 
as in Figure \ref{fig:f2}.}
\label{fig:f7}
\end{figure}

\begin{figure*}
  \centering
  \includegraphics[width= \linewidth,clip=true]{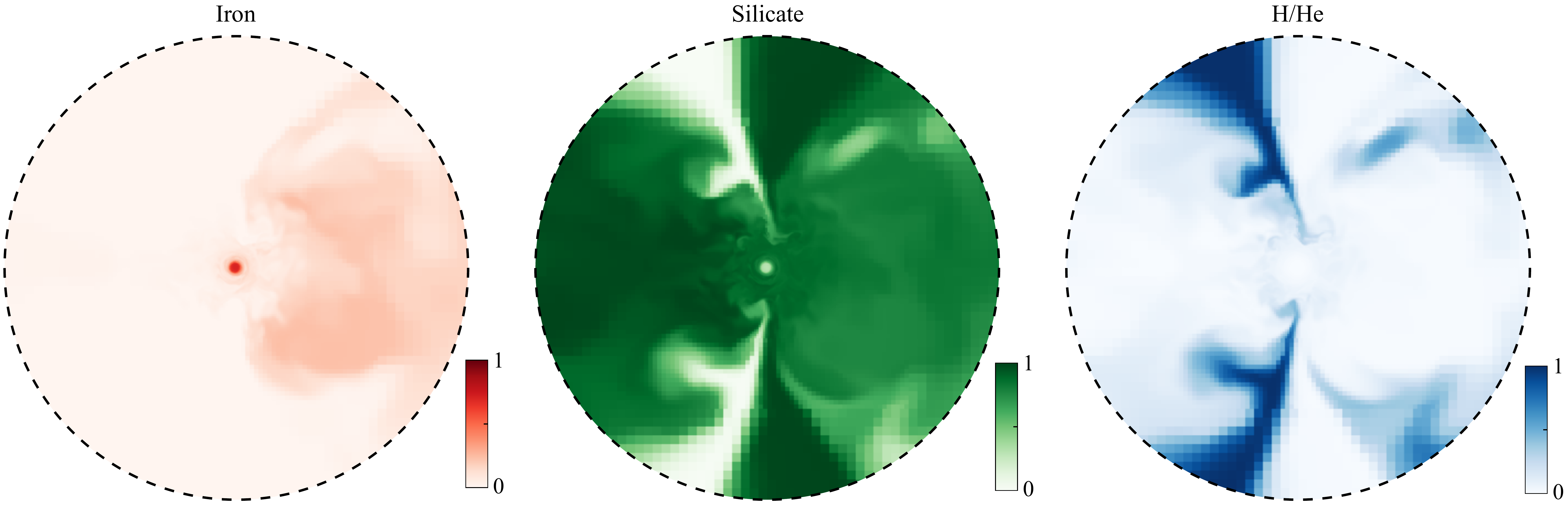}
\caption{The mass fraction slices of each species through the orbital plane at 21.5 hours after the high-speed impact. Mixing becomes less efficient because most of the H/He envelope has been blown away by the impactor. Dashed circles represent the Hill sphere with a radius of $3.21 \times 10^{10}\,{\rm cm}$.
}
\label{fig:f8}
\end{figure*}

In contrast, the rocky material is dredged up to the top of the 
atmosphere due to the turbulent motions driven by the impact.
As shown in Figure \ref{fig:f5}, we note that the target's 
interior, i.e. within $6 \times 10^9\,{\rm cm}$, has a steep and 
positive compositional gradient, $\nabla_Z = {\rm d}\log Z/{\rm d}\log P$, 
which serves to stabilize thermal convection in the mantle. At least 
right after the giant impacts, a head-on collision develops a hot and 
inhomogeneous interior. Heat transport deep inside a post-impact 
planet may be suppressed by the effect of double diffusion convection 
\citep[e.g.][]{2011ApJ...731...66R, 2012A&A...540A..20L}. Such an inefficient heat 
transfer mechanism would also prolong the presence of a magma ocean 
and delay the differentiation of iron material from a silicate-rich mantle.
Off-axis collisions would lead to more efficient mixing, but also a less
efficient kinetic energy contribution.

We summarize the energy budget in Table \ref{tab:t1}. Note that the 
planet's internal energy is almost doubled after the impact and 
it is larger than the sum of the internal and kinetic energies of 
the target and the impactor prior to the impact. This difference arises 
because the collision is accretionary, i.e. a large amount of the
impactor's mass is delivered to the interior of the target. As 
a result, not only most of the kinetic energy but also some portion 
of the gravitational potential energy is converted into the post-impact
internal energy. Therefore, one would expect that the temperature of 
the planet's interior to increase substantially after the impact.

\subsection{High-speed model 2}
The consequences of a high-speed impact is similar to that of the 
low-speed counterpart in terms of shock propagation throughout 
the target after the impact, mass loss via Roche lobe overflow, 
and gradually decaying oscillations inside the planet (Figure \ref{fig:f6}). Despite 
those general similarities, the high-speed impact differs from 
the low-speed one in three respects.

First, the high-speed impact is erosive. The mass of the target 
planet at 21.5 hr after the impact is about $9.9 M_\earth$, 
i.e., $0.1 M_\earth$ is stripped off from the target by the impactor. 
Thus, the total mass loss is $\sim 1.1 \, M_\oplus$. Besides, a 
significant amount of mass is levitated at the top of the atmosphere 
(see Figure \ref{fig:f7}). A breakdown of the mass budget in the 
Hill sphere (i.e., 3.21$\times 10^{10}$ cm in this case) shows that 
the impactor successfully deposits all of its iron material into the target, 
while most of the impactor's rocky material is ejected (see also Table 
\ref{tab:t1}). And the target retains about $1/5$ of its original 
atmosphere after such an energetic impact. 

Second, silicate rock material become the dominant species in the 
planet's atmosphere (Figure \ref{fig:f7} and \ref{fig:f8}). 
As a result, the compositional gradient is nearly zero out to the Hill's
radius, indicating a homogeneous distribution of the heavy material 
inside the bulk of the target (see Figure \ref{fig:f5}). In contrast
to the low-velocity model, model 1, the negligible compositional gradient 
cannot suppress convection in the interior of the target planet. 
Nevertheless, iron material remains concentrated near the center of
the planet as in the case of the low-speed model, model 1. 

Third, the planet becomes less bound gravitationally. The mass contained 
within the target's original size is much smaller (see Figure \ref{fig:f7}). 
In terms of the energy budget, most of the kinetic energy is 
carried away by the unbound mass and only a fraction of it
is directly converted to the planet's internal energy. However, the 
gravitational potential energy released by the sedimentation of 
rock material in the atmosphere becomes an extra heat 
source for the target interior. 
Assuming that silicates in the H/He atmosphere establish a local thermal 
equilibrium with the turbulent gas, they condense into grains in 
cooler outer regions with $r \gtrsim 10^{10}\,{\rm cm}$. These silicate 
condensates grow through collisional coagulation
\citep[e.g.][]{2003Icar..165..428P}. Relatively large silicate grains are 
expected to settle toward the planetary surface and sublimate along the 
way.

\section{Discussion and Conclusions}

The results presented in the previous sections show the planet's structure
immediately after head-on giant impacts. Models 1 and 2 illustrate the possibility 
of diverse instantaneous outcomes depending on energy of collision, one at low velocity and
the other at approximately the local Keplerian velocity, three times the planets' escape
velocity.  Whereas the initial gaseous envelope of a target 
super-Earth is mostly retained during the low-velocity impact, it is severely
depleted shortly after the high-velocity impact. 
In this context, we suggest that giant impacts of varying energy can effectively 
devolatilize super-Earths/mini-Neptunes, and can diversify the 
compositions and interior structures of these mid-sized extrasolar planets,
perhaps as occurred in our terrestrial system \citep[see, e.g.][]{2014AREPS..42..551A}.

The interior and atmosphere of a post-impact planet can achieve 
a thermal equilibrium state after a long-term evolution, which is difficult 
and not cost-effective to study by performing hydrodynamic simulations. 
Alternatively, the radiative cooling timescale ($\tau_{\rm rad}$) can be estimated from
the rate of temperature change in an atmospheric layer via the outgoing 
infrared radiation:
$\tau_{\rm rad} \sim P\,C_{\rm p}/(4 g \sigma T^3)$, where $P$ and $T$ 
are the pressure and temperature near the photosphere obtained from the end state of
hydrodynamic simulations, $g$ is the surface 
gravity, and $C_{\rm p}$ is the specific heat capacity at a 
constant pressure, and $\sigma$ is the Stefan-Boltzmann constant.
Under the intense radiation from a $1\,M_\odot$ central star,
an isothermal layer develops in the planet's residual upper atmosphere.

At the current location of Kepler 36b and c, the equilibrium 
temperature due to stellar irradiation is $\sim 900\,{\rm K}$.
Adopting EOS of an ideal gas for H/He and the gas opacity 
of \citet{2008ApJS..174..504F}, we find that the H/He gas within 
the Hill sphere is optically thick.  At the photosphere, 
we used $P \sim 2g/3\kappa$ and $\kappa$ is the opacity to estimate
$\tau_{\rm rad}$ to be larger than several days.  This estimate
justifies, {\it a posteriori} the neglect of stellar irradiation in our
impact simulations (see \S2). Since this cooling time scale is much
longer than the dynamical time scale, we can also assume that after
a giant impact, a heated planet evolves adiabatically into a hydrostatic 
equilibrium.

During the post-impact phase, the planet's atmosphere and interior contract as
the planet cools. Above the photosphere, a highly-irradiated planet has an isothermal layer.
We consider that an inflated planet after an impact radiates away heat from the photosphere.
\citet{2015arXiv150602049O} estimated the Kelvin-Helmholtz timescale of a planet (3, 5, and 10\,$M_\oplus$) with the equilibrium temperature of 900\,K 
as a function of planetary radius in unit of the Bondi radius and envelope mass fraction based on a stellar evolution code, i.e., MESA \citep{2011ApJS..192....3P}. 
We applied their results of 5 and 10\,$M_\oplus$ shown in Figure\,2 in \citet{2015arXiv150602049O} to our targets after giant impacts
($\sim 4.991\,M_\oplus$ and $9.738\,M_\oplus$ with the equilibrium temperature of $\sim 900$ K).

Following results of \citet{2015arXiv150602049O}, we 
estimate the timescale of the Kelvin-Helmholtz contraction to be $\sim 1$ Myr 
for the low-speed case and $< 10$ kyr for the high-speed one. A protracted 
state of a hot and inflated atmosphere enhances its mass loss via a Parker 
wind.  Based on Eq.(16) in \citet{2015arXiv150602049O}, we estimate that 
the amount of mass loss for post-impact planets via a Parker wind for the 
two cases would be $\sim 88 \%$ and $\sim 80 \%$ of its envelope mass 
right after the impact.

As the planet cools down, silicate material sediments from its atmosphere 
and contracts within its interior. The release of gravitational energy 
provides a source of internal heat which prolongs the inflationary state 
of the planet's atmosphere and enhances mass loss via a Parker wind.
The extended planetary atmosphere also intercepts a greater fraction of 
the stellar X-ray and UV irradiation and increases the rate of mass loss
through photoevaporation.

We estimate the mass loss rate of an extended atmosphere via stellar XUV irradiation after giant impacts.
Adopting analytical descriptions of a XUV energy flux from solar-type stars \citep{2005ApJ...622..680R},
the incident XUV luminosity at the Hill radius is $1.0 \times 10^{-7}\,L_\odot$ and $1.5 \times 10^{-7}\,L_\odot$ 
in the low-speed and high-speed model, respectively. 
Given that the conversion efficiency from XUV photons to the kinetic energy of bulk atmospheric outflow
is 10\% as a fiducial value \citep{2004Icar..170..167Y}, 
the mass loss rate from the Roche lobe via stellar XUV irradiation right after giant impacts
is estimated to be $\sim 3\,M_\oplus\,{\rm Myr}^{-1}$ for the former and $\sim 2\,M_\oplus\,{\rm Myr}^{-1}$ for the latter.
If we consider a typical decay timescale of a XUV flux for Sun-like stars is $\sim 0.1$Gyr, 
the planet in the low-speed model may lose the entire envelope because its atmosphere shrinks in a few Myr.
On the other hands, the planet in the high-speed model can contract more quickly ($< 10\,{\rm kyr}$), but
it is unlikely for its tenuous post-impact atmosphere to survive a subsequent mass loss via stellar irradiation, either. Note that the stellar tidal field is crucial in this context because it can continuously remove planetary outer atmosphere that is beyond the Hill sphere. For giant impacts happen further away from the host star, tidal stripping becomes inefficient. And the outcome of a giant impact is determined by the mass ratio, impact speed and impact angle \citep{2010ChEG...70..199A,2012ApJ...745...79L,2012ApJ...751...32S}.

To summarize, severe giant impacts can significantly devolatilize close-in super-Earths and mini-Neptunes. The compositional dichotomy between Kepler-36 b and c can be explained by their distinct impact histories along with the formation of the closely packed system, i.e. Kepler-36b experienced substantial giant impacts, and in the meanwhile Kepler-36c survived from being heavily bombarded. In addition, we speculate that giant impacts may have been imprinted in the large dispersion in the mass--radius relationship of close-in sub-Neptune-sized planets, as giant impacts occur stochastically and can diversify planetary interior and atmosphere otherwise well constrained.

In our solar system, there may be evidence for a similar though more
subtle dichotomy. Despite their similar masses, radii, and compositions, thermal evolution 
models of Uranus and Neptune suggest that right after their formation, 
Neptune may have been relatively luminous but Uranus relatively 
faint \citep[e.g.][]{1980JGR....85..225H, 2011ApJ...729...32F}. 
\citet{1986LPI....17.1011S} suggested that this dichotomy may be accounted
for if a violent head-on collision yielded a hot and homogeneous interior 
of Neptune, whereas an oblique collision caused a tilted Uranus with a 
stably-stratified interior. The results presented here imply that the 
former scenario for Neptune may be possible.

\acknowledgements

We thank Jonathan Fortney, Pascale Garaud, James Guillochon, Don Korycansky, and 
Xiaojia Zhang for fruitful discussions. We also thank an anonymous 
referee for helpful suggestions to improve the clarity of this manuscript. 
S.-F.L. and E.A. are sponsored by the 
NASA grant NNX13AR66G. Y.H. is supported by a Grant-in-Aid for JSPS 
Fellows (No.25000465) and a Grant-in-Aid for Scientific Research on 
Innovative Areas (No. 26103711) from MEXT of Japan. Support from a 
UC/Lab Fee grant, an IGPPS grant, and the NSF grant 1211394 are also acknowledged. Numerical 
computations were carried out on the Laohu cluster at NAOC, on Cray 
XC30 at NAOJ, and on the Hyades cluster at UCSC. The FLASH code used in this work was in part developed by the DOE NNSA-ASC OASCR Flash Center at the University of Chicago. VisIt is supported by the Department of Energy with funding from the Advanced Simulation and Computing Program and the Scientific Discovery through Advanced Computing Program.

\end{CJK*}

\end{document}